\documentclass[12pt]{iopart}
\usepackage{iopams}
\usepackage{epsfig}   
\usepackage{graphics}
\newcommand{\apj}{Astrophys. J.}
\newcommand{\apjl}{Astrophys. J. Lett.}
\newcommand{\aj}{Astron. J.}
\newcommand{\aap}{Astron. Astrophys.}
\newcommand{\prd}{Phys. Rev. D}
\newcommand{\mnras}{Mon. Not. R. Astron. Soc.}
\begin{document} 

\title[Probing dark energy inhomogeneities with supernovae]
{Probing dark energy inhomogeneities with supernovae}

\author{Michael Blomqvist$^1$, Edvard M\"ortsell$^2$ and Serena
Nobili$^2$}

\address{$^1$ Department of Astronomy, Stockholm University, AlbaNova
         University Center \\
         S--106 91 Stockholm, Sweden}
\address{$^2$ Department of Physics, Stockholm University, AlbaNova
         University Center \\ S--106 91 Stockholm, Sweden}

\ead{\mailto{michaelb@astro.su.se}, \mailto{edvard@physto.se} and
\mailto{serena@physto.se}}

\begin{abstract}
We discuss the possibility to identify anisotropic and/or inhomogeneous 
cosmological models using type Ia supernova data. A search for correlations 
in current type Ia peak magnitudes over a large range of angular
scales yields a null result.  However, the same analysis limited to
supernovae at low redshift, shows a feeble anticorrelation at the
2$\sigma$ level at angular scales $\theta \approx 40^{\circ}$. 
Upcoming data from, e.g., the SNLS (Supernova Legacy Survey) and the SDSS-II
(SDSS: Sloan Digital Sky Survey) supernova searches 
will improve our limits on the size of -- or possibly detect --
possible correlations also at high redshift at the per cent level  
in the near future. With data from the proposed SNAP (SuperNova Acceleration Probe)
satellite, we will be able to detect the induced correlations from gravitational
lensing on type Ia peak magnitudes on scales less than a degree.

\end{abstract}

\noindent{\it Keywords}: dark energy theory, supernova type Ia
%\maketitle 

%%%%%%%%%%%%%%%%%%%%%%%%%%%%%%%%%%%%%%%%%%%%%%%%%%%%%%%%%%%%%%%%%%%%%%%
\section{Introduction}
%%%%%%%%%%%%%%%%%%%%%%%%%%%%%%%%%%%%%%%%%%%%%%%%%%%%%%%%%%%%%%%%%%%%%%%
Apart from the identification of dark matter, the most dominant question
in cosmology today is that of what is responsible for the apparent
acceleration of the universal expansion. Assuming that the cause is a
dominant energy component with negative pressure, current efforts are
focused on establishing whether this dark energy (DE) component can be described using a
cosmological constant (CC), i.e. an energy component with constant
density and a fixed equation of state (EOS) given by $p = \omega\rho$
where $w=-1$, or whether dark energy is dynamical (DDE) with a varying
EOS, $w=w(z)$.

Two main techniques are employed in these investigations: probing
cosmological distances and probing the growth of cosmological
structures, both of which are sensitive to the energy content of the
universe. Distances are most robustly probed via type Ia supernovae (SNe~Ia)
\cite{1998AJ....116.1009R,1999ApJ...517..565P,2006A&A...447...31A,2007ApJ...666..694W},
the baryon acoustic oscillation (BAO) peak \cite{2005MNRAS.362..505C,2005ApJ...633..560E,2007MNRAS.381.1053P}  
and the last scattering surface of the cosmic microwave background (CMB) \cite{2008arXiv0803.0547K}.  
Structure growth is mainly probed by large galaxy surveys such as the 2dF \cite{2003MNRAS.346...78H} and the SDSS \cite{2004ApJ...607..655P,2007ApJ...657..645P} but
also via galaxy cluster counts \cite{2006ApJ...653..954D} and weak lensing \cite{2006ApJ...647..116H}. Despite some claims to
the contrary, there is a consensus that current data are perfectly
consistent with having approximately 70\% of the energy density in
the universe in the form of a CC \cite{2007ApJ...666..716D}. 

However, cosmological distances only depend on the dark energy
EOS in the form of a double integral and it is thus very difficult to
investigate the time evolution of the EOS parameter, $w(z)$. Though
not demanding it, current data thus still allow for large deviations
of $w(z)$ from the CC value of $w=-1$ \cite{2007ApJ...666..716D}. Even if the limits on $w(z)$ will improve in the future, detecting such an evolution remains notoriously
difficult, especially if the deviation from the CC value turns out to be small.

It is thus of interest to look for alternative ways to detect or
constrain any behaviour of the DE that differs from that of a
CC. One such possibility is to study not temporal evolution, but rather
spatial variations in DE properties. The CC value $w=-1$ is special in
the sense that it not only gives a constant energy density in time,
but also that it prevents any spatial clustering of DE. Any deviations
from this value inevitably cause the DE to cluster, a clustering which --
if detected -- would refute the CC as the dominant energy
component in the universe.

The clustering properties of DDE in different models and scenarios
have been investigated in several recent articles (see, e.g., \cite{2007arXiv0709.2227M} and references therein). For values close to $w=-1$, such clustering is expected to be weak and take place mainly on very large scales, larger than the current Hubble radius and
thus inaccessible to observations. On smaller scales, it can be shown
that the DE density will tend to be anticorrelated with the matter
density \cite{2007arXiv0709.2227M,2007PhRvD..75f3507D}. Although,
again, the amount of clustering is expected to be small (on the order
of $10^{-5}[1+w]$), it is nevertheless important to get observational
confirmation that this is indeed the case, even if it is not expected
from purely theoretical considerations. On a more speculative note,
also in models where the apparent acceleration of the universe is explained
in terms of the back-reaction of inhomogeneities in the universe,
we expect to have spatial variations in, e.g., cosmological distances
\cite{2008PhRvD..77b3003M}. Furthermore, cosmological observations can be
used to put constraints on the degree of centricity in spherically
symmetric inhomogeneous models
\cite{2007JCAP...02...19E,2007PhRvD..75b3506A} and on anisotropic DE
models
\cite{2007arXiv0707.0279K,2006MNRAS.371.1373M,2008PhRvD..77b3534R,2007MNRAS.382..793M,2006PhRvD..74d3505T}. 

One way to probe models with inhomogeneous DE or back-reaction is
to look for anisotropies in the observed peak magnitudes of SNe~Ia. 
In general, we expect cosmological distances 
to be correlated on scales similar to the clustering scale of the DE or the 
matter inhomogeneities responsible for the back-reaction effect.
The detection of such correlations is demanding because of the
intrinsic variation and observational uncertainties in SN~Ia peak magnitudes. 

Since systematic effects connected to, e.g., the 
observational properties of different telescopes, calibration issues and details of 
the light-curve fitting procedure can induce correlations in SN~Ia magnitudes, it is 
important to minimize these effects by using as homogeneous a data set as possible 
in the analysis.
Also, physical effects such as peculiar motions
\cite{2007ApJ...661..650H}, gravitational lensing \cite{2006PhRvL..96b1301C} and dust 
extinction \cite{2007ApJ...657...71Z} will introduce correlations in the SN Ia data and 
need to be controlled and, if possible, corrected for.  

In Kolatt \& Lahav \cite{2001MNRAS.323..859K}, an early data set 
consisting of a total of 79 SNe~Ia combined from the Supernova Cosmology Project 
\cite{1999ApJ...517..565P} and the High-$z$ Supernova Search Team 
\cite{1998AJ....116.1009R}  was used to look for directional variations 
in the best fit cosmological parameters. The result is consistent with the expected 
statistical variations in a homogeneous universe.  
Gupta \textit{et al} \cite{2007astro.ph..1683G} used a more homogeneous
set of 157 SNe~Ia \cite{2004ApJ...607..665R} to look for directional
variations in the statistical scatter around the best fit cosmological
model, again with a result consistent with null variations. A slightly
different technique was used in Bochner
\cite{2007astro.ph..2730B} to look for anisotropic scatter in 172
SNe~Ia \cite{2003ApJ...594....1T} with similar results. Although
current investigations have been mainly inconclusive, Schwartz \&
Weinhorst \cite{2007A&A...474..717S} find an offset in the
best fit calibration of low $z$ SNe~Ia, between the north and south
equatorial hemispheres at the 95\% confidence level. Whether this
hint of a north/south asymmetry is due to a statistical coincidence,
observational systematic differences, or correlated peculiar motions or
has a cosmological origin is not clear at the moment. 

In this paper we devise a general methodology for detecting angular
correlations in SN~Ia magnitude residuals. We apply this method to current
SN~Ia data as well as simulated upcoming data sets.

In section~\ref{method}, we describe the method for analysing SN~Ia
magnitude residuals for angular correlations and define the detection
limit for such correlations. Two data sets from the literature are analysed 
for angular correlations in section~\ref{analyse}. In section~\ref{future}, we present 
different SN Ia surveys and investigate the detection limits for each of
them. In section~\ref{toymodel}, we assume a toy model correlation function
and illustrate how future data can put limits on such a
correlation. The paper is concluded in section~\ref{conclusions}. 

%%%%%%%%%%%%%%%%%%%%%%%%%%%%%%%%%%%%%%%%%%%%%%%%%%%%%%%%%%%%%%%%%%%%%%%
\section{Method}\label{method}
%%%%%%%%%%%%%%%%%%%%%%%%%%%%%%%%%%%%%%%%%%%%%%%%%%%%%%%%%%%%%%%%%%%%%%%

In this paper we study angular correlations only, since
the effect from possible DE inhomogeneities is larger in the
transverse direction than in the radial direction. Moreover, a full
three-dimensional study would require larger statistics than
currently available. The data sets we investigate contain both nearby
and distant SNe~Ia, for which we have redshifts, $z$, observed peak
magnitudes, $m_{\rm obs}$, with uncertainties, $\sigma _m$, as well as
positions on the sky, right ascension, $\alpha$, and declination,
$\delta$. 

%======================================================================
\subsection{The angular correlation function}
\label{angcorr}
%======================================================================
The magnitude residual of a SN~Ia on the Hubble diagram is the
difference between its observed magnitude and the magnitude predicted
from its redshift in the best fit cosmology,
\begin{equation}
\delta m =m_{\rm obs} - m_{\rm fit}\ .
\end{equation}
We define a normalized and scaled residual, $X$, by subtracting the mean
and dividing by the uncertainty,
\begin{equation}
X\equiv \frac{\delta m - \mu_{\delta m}}{\sigma _m}\ .
\end{equation}
Note that the $\sigma _m$ include both observational errors and the
intrinsic dispersion of the SN~Ia peak magnitudes. Thus, for uncorrelated
Gaussian errors, we expect $X$ to follow a normal distribution
with mean $\mu_{X}\approx 0$ and dispersion $\sigma _{X}\approx 1$.

The angular correlation function of the magnitude residuals 
for a certain angular separation is defined as the expectation
value of the product of the magnitude residuals of all SN pairs
separated by that angle,
\begin{equation}\label{corrfun}
C(\theta)=\langle X\cdot X(\theta)\rangle\ ,
\end{equation} 
where the angular separation, $\theta$, between two SNe (denoted 1 and 2) is
given by
\begin{equation}
\cos (\theta) =\sin(90^{\circ}-\delta _{1}) \sin(90^{\circ}-\delta _{2}
)\cos (\alpha _1 - \alpha _2)+\cos(90^{\circ}-\delta
_{1})\cos(90^{\circ}-\delta _{2})\ .
\end{equation} 
Note that our definition of the correlation function is equivalent to the
correlation coefficient for the original residuals, $\delta m$,
\begin{equation}
C(\theta)\equiv \rho_{\delta m \delta m(\theta)}\ .
\end{equation} 
If $C(\theta)>0$, the magnitude residuals are correlated,
if $C(\theta)=0$ they are uncorrelated and if $C(\theta)<0$ they are
anticorrelated. 

%======================================================================
\subsection{Detection limits for Gaussian magnitude
  residuals}
\label{detect}
%======================================================================
Under the assumption that the magnitude residuals follow a Gaussian distribution
with a mean of zero and a variance $\sigma _{X}^{2}\approx 1$, their
product will follow a probability density distribution given by
\begin{equation}
p(X\cdot X)=\frac{1}{\pi \sigma _{X}^{2}}K_{0}\Bigg(\frac{\vert X\cdot
X\vert}{\sigma _{X}^{2}} \Bigg)\ ,
\end{equation}
where $K_{0}$ is the modified Bessel function of the second kind and $X\cdot X$ 
denotes all possible products between the magnitude residuals. The
variance of this probability density distribution is $\sigma _{X\cdot
X}^{2}=\sigma _{X}^{4}$. 

The correlation for a given angular separation is the expectation
value of the product of the magnitude residuals for $N_{\rm p}$
SN pairs, where $N_{\rm p}$ is the number of unique pairs that can be
defined from a set of $N$ SNe,
\begin{equation}
N_{\rm p}=\frac{N^2-N}{2}\ .
\end{equation}
The dispersion in this expectation value is given by
\begin{equation}
\sigma_C=\frac{\sigma _{X\cdot X}}{\sqrt N_{\rm p}}=\frac{\sigma
_{X}^{2}}{\sqrt N_{\rm p}}\ .
\end{equation}
Since $\sigma _{X}^{2}\approx 1$, we can define the $n$-standard deviation ($n$-$\sigma$)
detection limit of the angular correlation to be
\begin{equation}\label{clim}
C_{\rm lim}=\frac{n}{\sqrt N_{\rm p}}\ .
\end{equation}

%%%%%%%%%%%%%%%%%%%%%%%%%%%%%%%%%%%%%%%%%%%%%%%%%%%%%%%%%%%%%%%%%%%%%%%
\section{Analysing current data for angular correlations}
\label{analyse}
%%%%%%%%%%%%%%%%%%%%%%%%%%%%%%%%%%%%%%%%%%%%%%%%%%%%%%%%%%%%%%%%%%%%%%%
In this section, we will apply the methodology described above to
two current SN~Ia data sets in order to search for possible angular
correlations in the magnitude residuals. We again stress the
importance of using as homogeneous a data set as possible in order to
minimize the impact of systematic effects connected to, e.g., the
light-curve fitting method used when deriving the peak magnitudes. 

%======================================================================
\subsection{Data sets}\label{datasets}
\begin{figure}
\begin{center}
\includegraphics[angle=0,width=.45\textwidth]{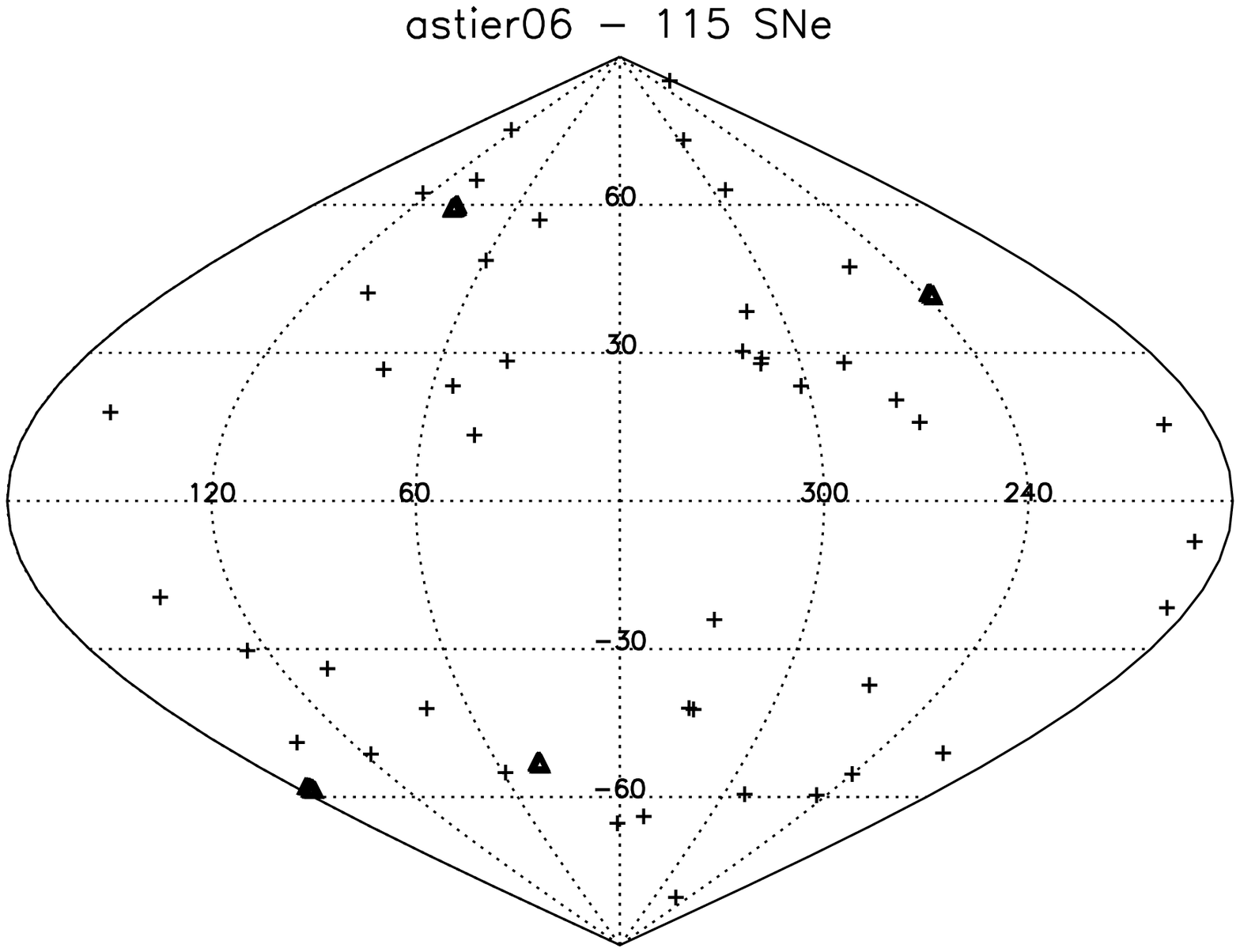}
\includegraphics[angle=0,width=.45\textwidth]{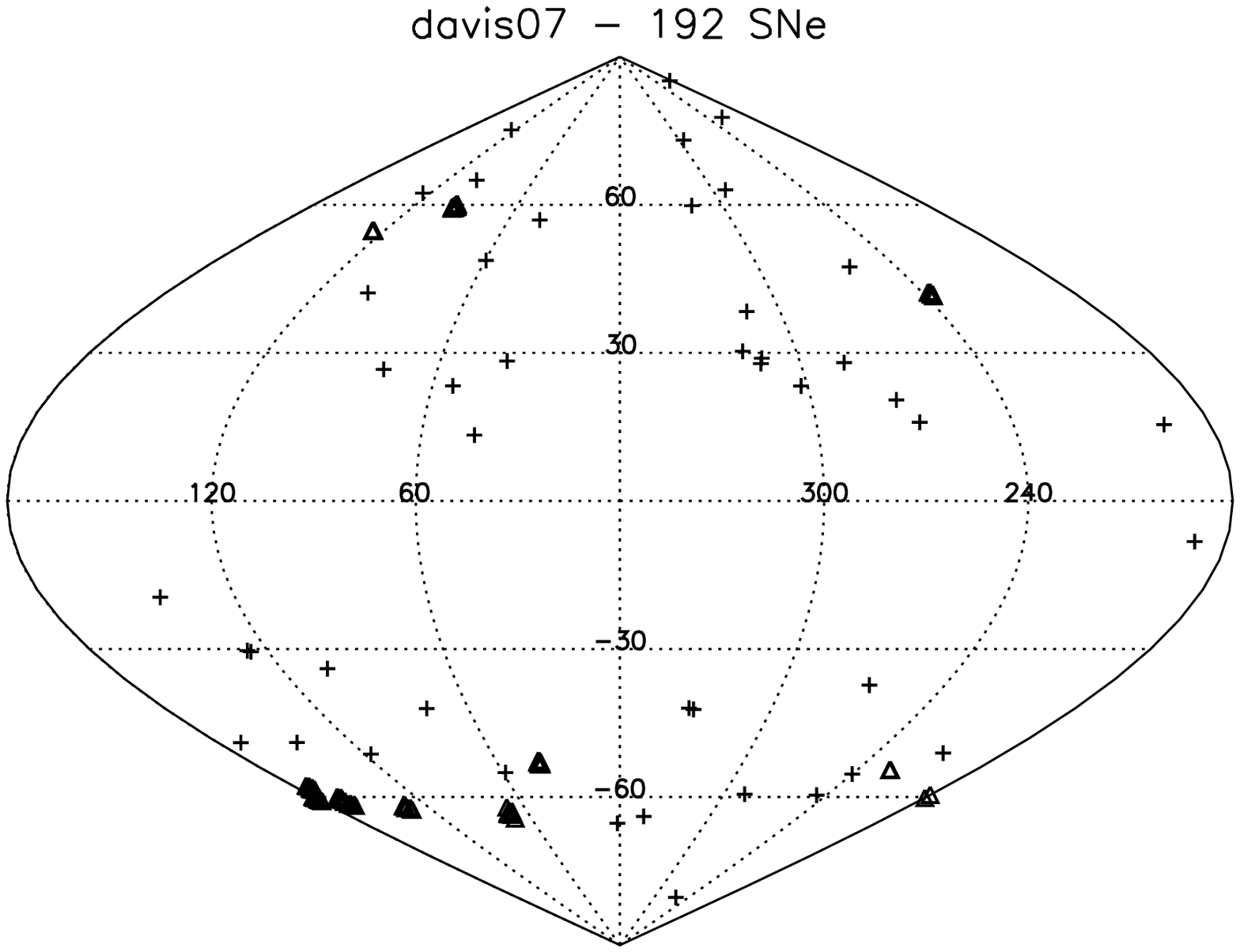}
\caption{\label{fig:snmaps} Distribution of SNe~Ia in the sky in
galactic coordinates. Left panel: 115 SNe~Ia in astier06, consisting
of 71 SNLS SNe (triangles) and 44 low $z$ SNe (plus signs). Note the
pencil beam geometry of the four SNLS survey patches. Right panel: 192
SNe~Ia in davis07, with 147 high $z$ SNe (triangles) and 45 low $z$ SNe (plus signs).
Many of the SNe are the same as in astier06.} 
\end{center}
\end{figure}

%======================================================================
The first year Supernova Legacy Survey (SNLS) data set in Astier \textit{et al}
\cite{2006A&A...447...31A} (we will denote this data set as astier06)
contains a total of 115 SNe~Ia, of which 71 are high $z$ SNLS SNe ($z>0.2$)
and 44 are nearby ones ($z<0.2$). Figure~\ref{fig:snmaps} (left panel)
shows the distribution of the SNe in the sky in galactic
coordinates. Whereas the nearby SNe~Ia (plus signs) are scattered across the
entire sky, the SNLS SNe (triangles) are confined to four pencil-beam patches,
each covering a one-square-degree area. These patches are referred to as
D1-D4 and contain 14, 8, 31 and 18 SNe, respectively. The SALT light-curve model was used 
to fit all the SN~Ia light curves in the data set. We add an intrinsic dispersion of $\sigma _{\rm int}=0.13$ to the uncertainties in the observed SN~Ia magnitudes provided.

The second data set is from Davis \textit{et al}
\cite{2006A&A...447...31A,2007ApJ...666..694W,2007ApJ...666..716D,2007ApJ...659...98R} 
(we denote this as davis07) and is a compilation of 192 SNe~Ia consisting
of 60 ESSENCE SNe, 57 SNLS SNe, 30 HST `gold' SNe and 45 nearby
SNe. Figure~\ref{fig:snmaps} (right panel) shows the distribution of
the SNe in the sky in galactic coordinates. Most of the low $z$ SNe
(plus signs) are the same as in astier06 and thus fairly equally
distributed across the sky. The 147 high $z$ SNe (diamonds) are, on
the other hand, very unevenly distributed. We use the peak magnitudes
and uncertainties reported by davis07, which were obtained using the
MCLS2k2 light-curve fitter on all photometric data. The uncertainties
include both the observational and the intrinsic magnitude scatter.

Since the two data sets have many SNe in common, e.g., most of the
nearby SNe, the corresponding correlation functions will not be
independent. Comparison of the results will provide a useful check on
the impact of the light-curve fitter used in deriving the SN~Ia peak
magnitudes on the angular correlation function.

%======================================================================
\subsection{Results}
%======================================================================
For a given data set, we find the best fit cosmological parameters, and
subtract the corresponding magnitudes, $m_{\rm fit}$, from the
observed magnitudes, $m_{\rm obs}$, to get the residuals, $\delta m$.
From these, we subtract the mean residual\footnote{The value of
  $\mu_{\delta m}$ is very close to zero.}, $\mu_{\delta m}$, and
divide by the uncertainty, $\sigma_m$, to obtain the normalized and
scaled magnitude residuals, $X$. 

The values of the magnitude residuals and, consequently, the
correlation function obviously depend to some degree on what cosmology
we subtract. However, by varying the cosmological parameters of the
subtracted model, we have verified that the impact on the correlation
function is very small.

\begin{figure}
\begin{center}
\includegraphics[angle=0,width=.48\textwidth]{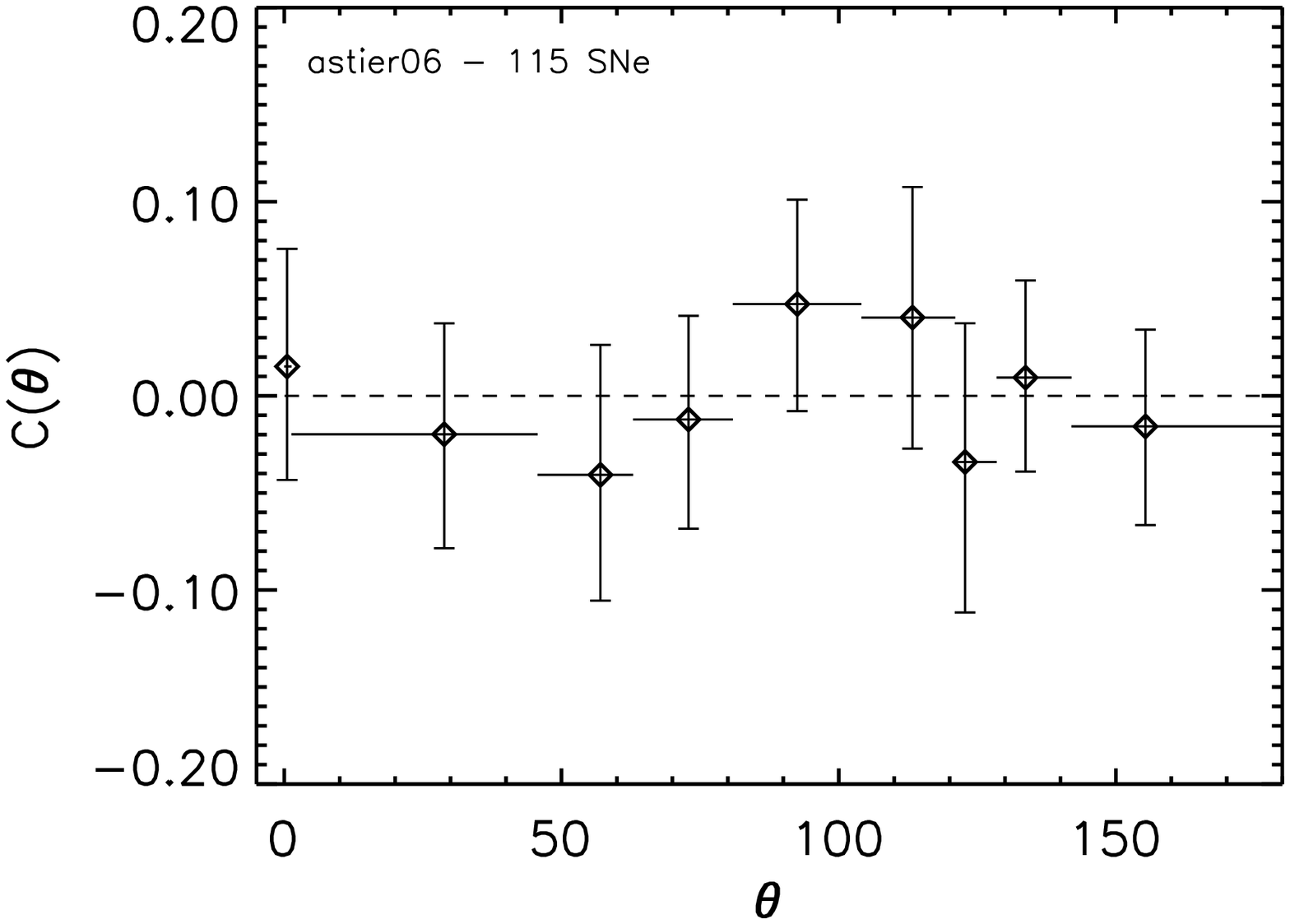}
\includegraphics[angle=0,width=.48\textwidth]{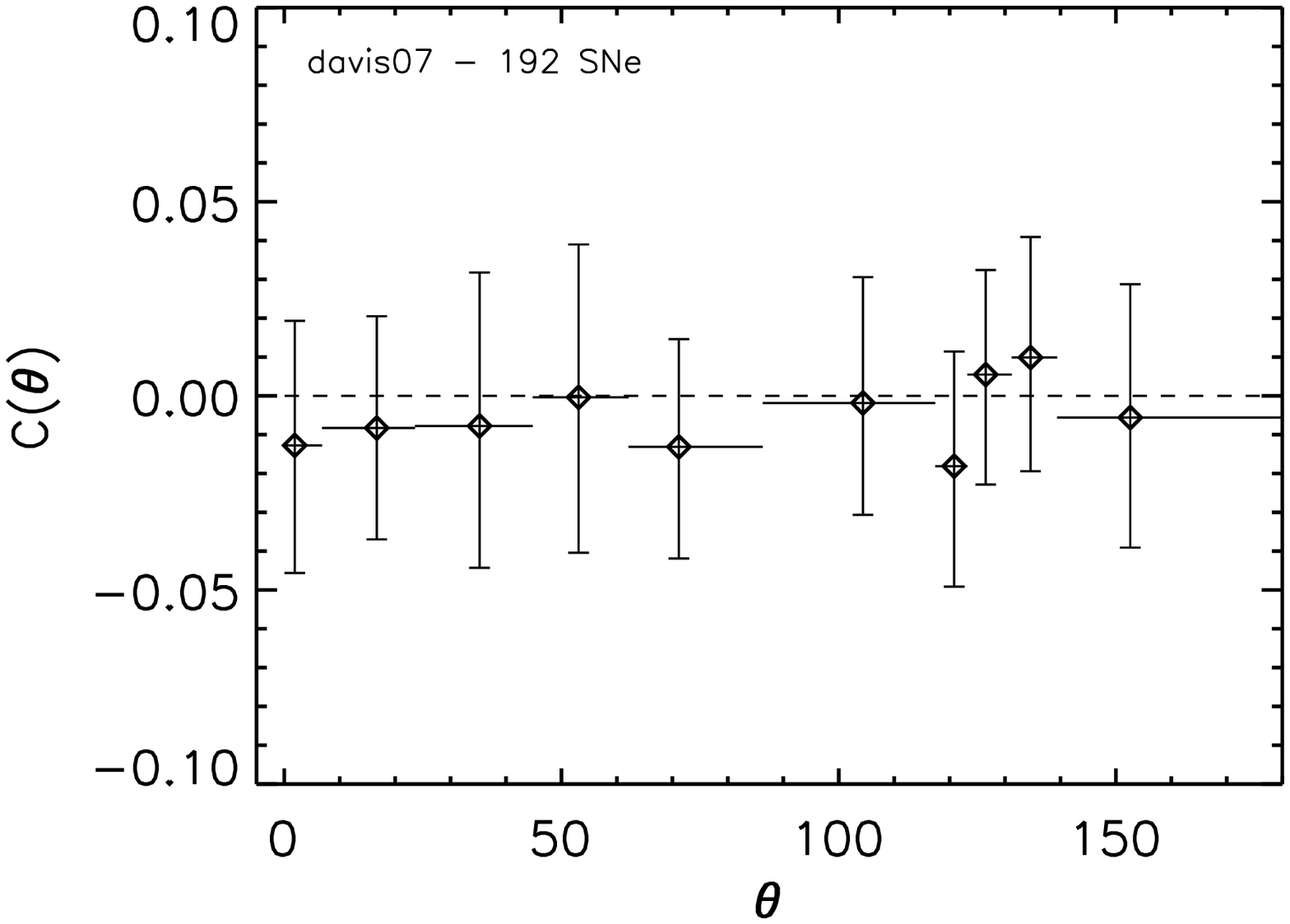}
\caption{\label{fig:astierdavis} Correlation functions for
the 115 SNe~Ia in astier06 and the 192 SNe~Ia in davis07.
The error bars represent the 68\% confidence limit. Each bin contains
approximately the same number of SN pairs. The horizontal bars
indicate the range of each bin and the points are placed at the
average angular separation in each bin. Note that, the data 
points are not independent, since each SN contributes to several bins.}   
\end{center}
\end{figure}

The correlation functions for the 115 SNe~Ia in astier06 and 192 SNe~Ia in
davis07 are presented in figure~\ref{fig:astierdavis}. The error bars
are obtained using Monte Carlo simulations where we generate 2000 new data
sets by adding random numbers from a Gaussian distribution with a mean
of zero and a variance of $\sigma _{X}^{2}=1$ to the magnitude
residuals of the SNe~Ia. These data sets are then analysed to
investigate the possible spread in the correlation function. The error
bars represent the 68\% confidence limit. The data is binned in such
a way that each bin contains approximately the same number of SN pairs
with an angular resolution of approximately $15^{\circ}$. Note that
since each SN~Ia is included in several bins, the data points will not
be independent from each other, as seen clearly in the right panel of
figure~\ref{fig:astierdavis}, where almost all the data are below the
line. Thus, we do not expect about 1/3 of the points to lie more than 1$\sigma$ away from the zero line, as would be the case for uncorrelated data points.
The horizontal bars indicate the range of each bin and the
points are placed at the average angular separation in each bin. Both
data sets are clearly consistent with the SN magnitude residuals being
uncorrelated, but allowing for correlations at the $\sim 10\%$ and 
$\sim 5\%$ level for the astier06 and davis07 samples respectively.

%======================================================================
\subsection{High $z$}
\begin{figure}
\begin{center}
\includegraphics[angle=0,width=.48\textwidth]{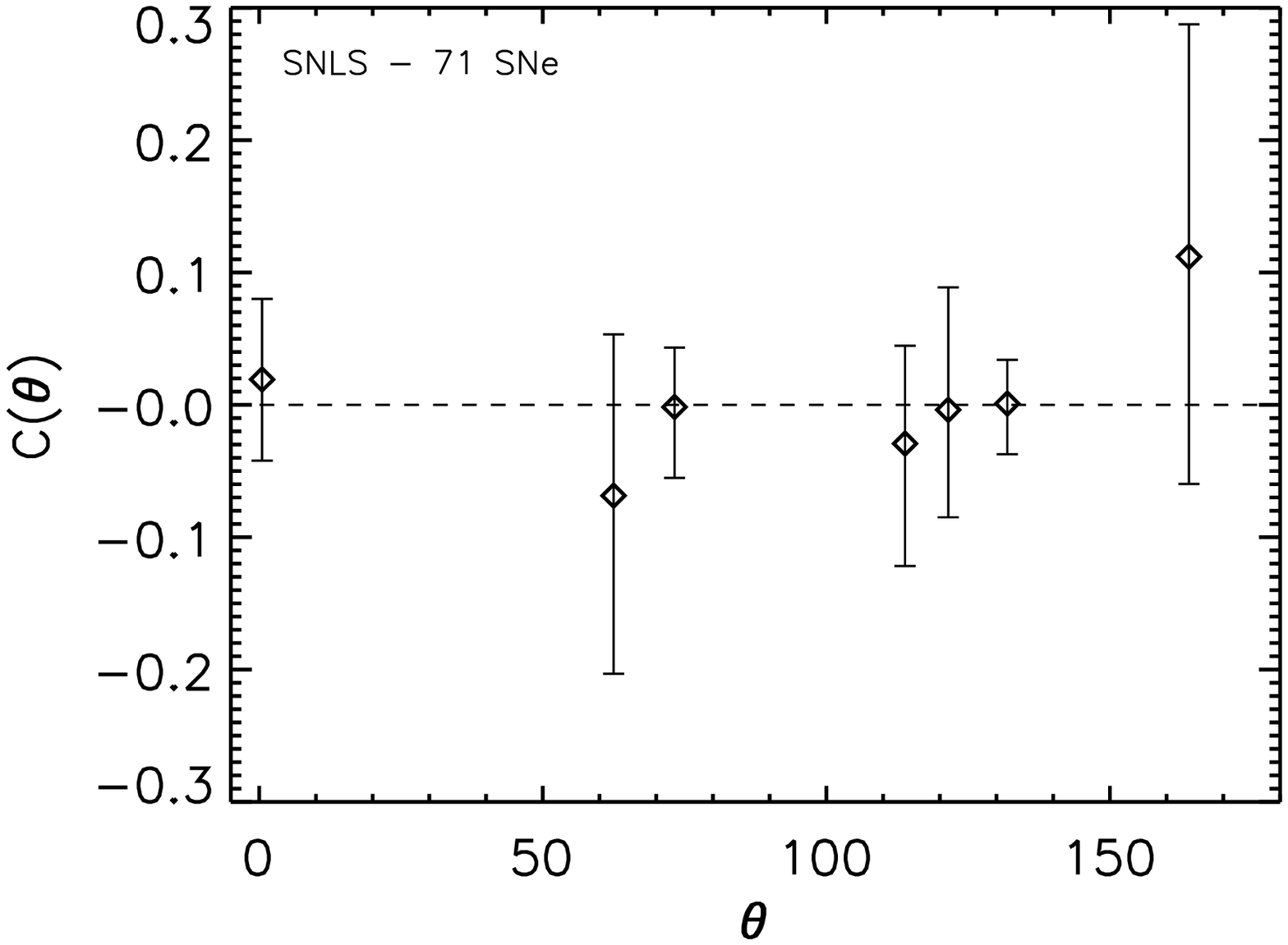}
\includegraphics[angle=0,width=.48\textwidth]{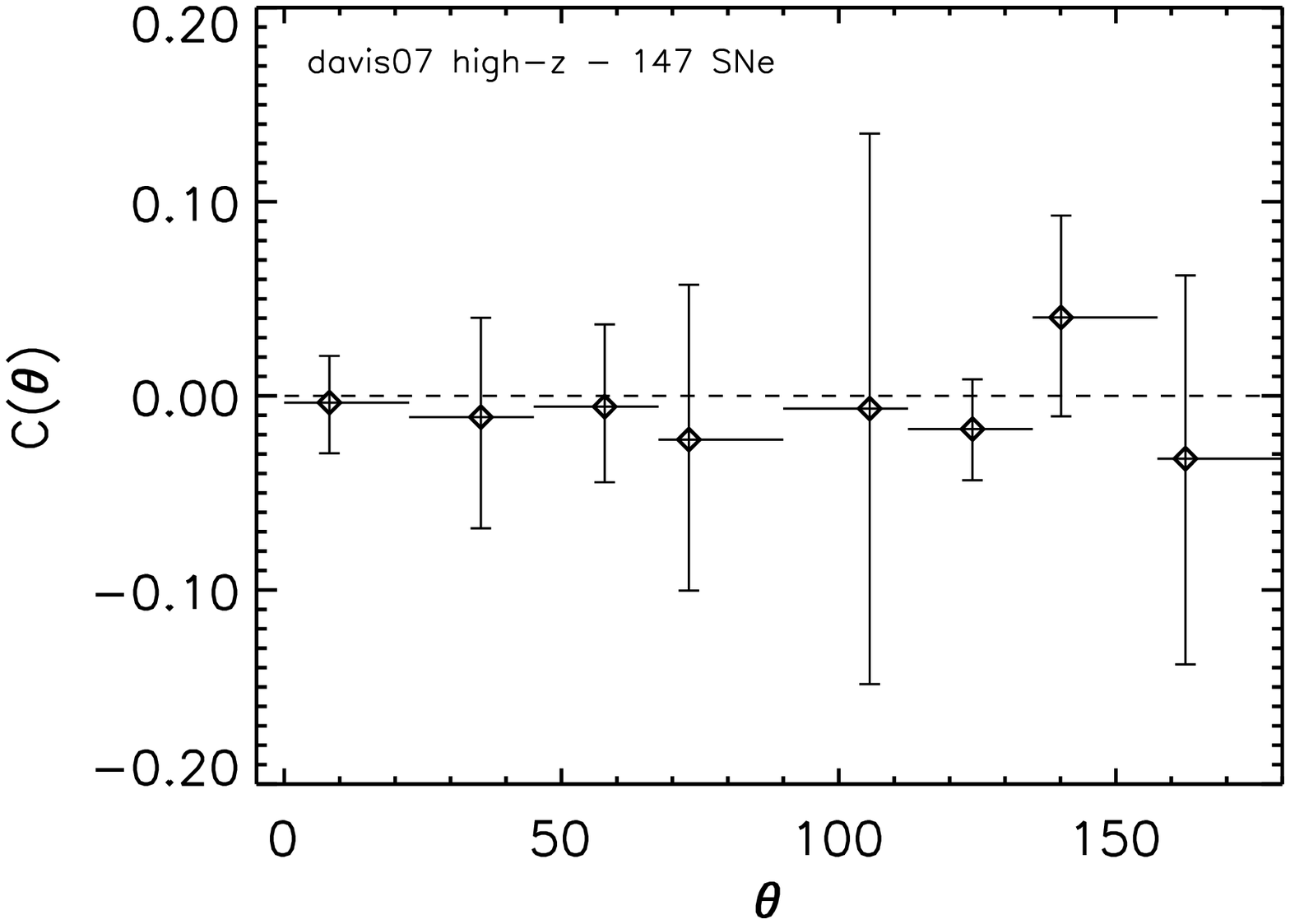}
\caption{\label{fig:highz} Correlation functions for the high $z$
  SNe~Ia in astier06 and davis07. The error bars represent the 68\%
  confidence limit. Left panel: correlation function for the 71 SNLS
  SNe in astier06. The bins are at the certain angular separations
  defined by the survey geometry. Right panel: correlation function
  for the 147 high $z$ SNe in davis07. The binning is done
  uniformly. The horizontal bars indicate the range of each bin and
  the points are placed at the average angular separation in each
  bin.} 
\end{center}
\end{figure}

%======================================================================
We expect the effects from inhomogeneous DE to be more prominent at
large distances and also 
systematic effects from, e.g., peculiar velocities to decrease with distance.
Therefore, we check the high $z$ SNe~Ia in both astier06 and davis07 for correlations
separately, i.e. we exclude the low $z$ SNe~Ia in the analysis. The
results are presented in figure~\ref{fig:highz}.  
The error bars represent the 68\% confidence limit and are obtained
using the same Monte Carlo simulations as above. 

The left panel in figure~\ref{fig:highz} shows the correlation
function for the 71 SNLS SNe in astier06. Due to the pencil-beam
geometry and the locations of the four patches of the SNLS survey, the
SNe will be separated only by certain angles. We therefore do the
binning at precisely these angles. The results are consistent with the
SN magnitude residuals being uncorrelated. We note that there are
differences in SN~Ia brightness between the patches, e.g., there is a
difference in the mean scaled magnitude residual between patch D1 and
D4 of $X \approx 0.65$
(where D4 is the brighter), corresponding to a difference of
$\langle\delta m\rangle\approx 0.11$ mag in the mean values of the
original residuals. However, this is consistent with the expected
statistical fluctuations at the 2$\sigma$ level. A search for
correlations inside the patches, i.e. on sub-degree scales, yields a
zero result for magnitude correlations also on small scales.  

The right panel in figure~\ref{fig:highz} shows the correlation function for the 
147 high $z$ SNe in davis07 adopting a resolution of $22.5^{\circ}$. 
Again, the results are consistent with the SN magnitude residuals being uncorrelated.

%======================================================================
\subsection{Low $z$}
\begin{figure}
\begin{center}
\includegraphics[angle=0,width=.48\textwidth]{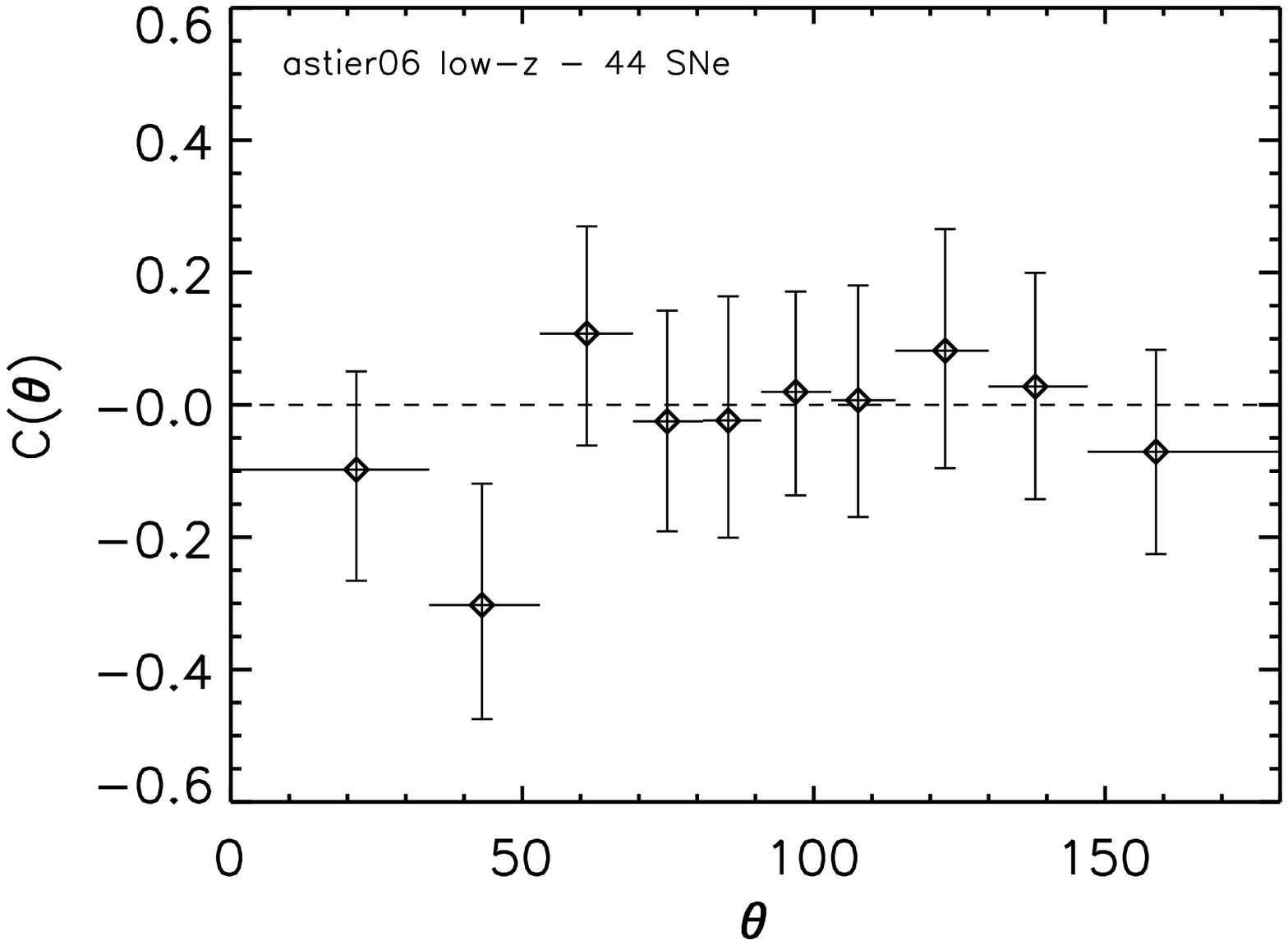}
\includegraphics[angle=0,width=.48\textwidth]{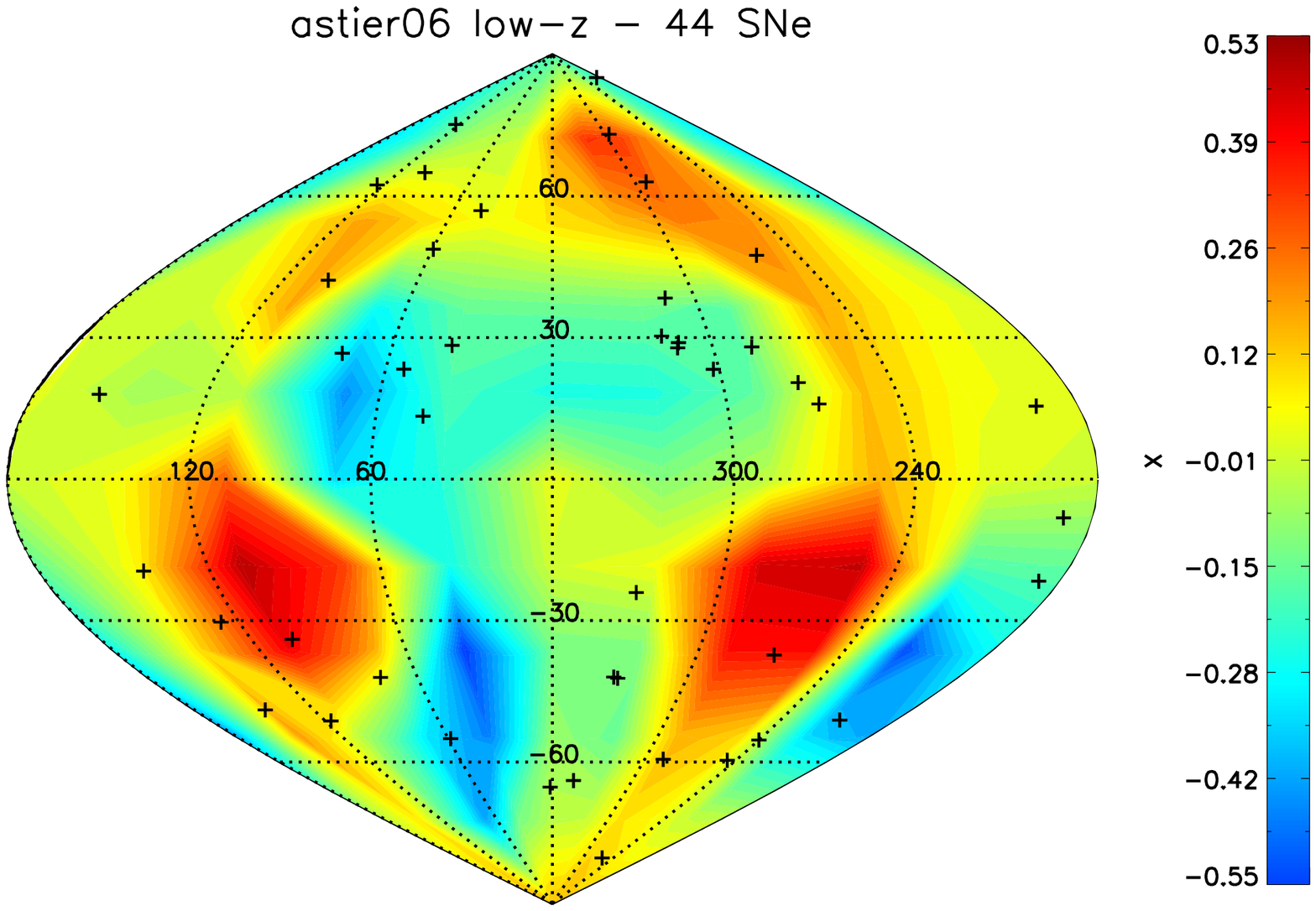}
\caption{\label{fig:lowz} Left panel: correlation function for the 44 low $z$ SNe in astier06. 
The error bars represent the 68\% confidence limit. Each bin contains approximately the same 
number of SN pairs. The horizontal bars indicate the range of each bin and the points are 
placed at the average angular separation in each bin. Right panel: colour map of the scaled magnitude residuals for the 44 low $z$ SNe in astier06 smoothed over an angular scale of $15^{\circ}$.}
\end{center}
\end{figure}

%======================================================================
We also analyze the low $z$ SNe, in the redshift range $0.015
\leq z \leq 0.125$, separately. 
As noted, correlations between SNe Ia at low redshifts could be due
to, e.g., correlations in their peculiar motions. However, since the
SNe~Ia in our sample are in the Hubble flow, we expect the
contribution to the peak magnitute to be small. 
The results for the 44 nearby SNe in astier06 are presented in
figure~\ref{fig:lowz} (left panel). The data are binned in such a way
that each bin contains approximately the same number of SN pairs with
an angular resolution of approximately $15^{\circ}$. We detect an
anticorrelation at the 2$\sigma$ level around $\theta \approx
40^{\circ}$. The 45 low $z$ SNe in davis07 (out of which 41 SNe are
also in astier06) show the same anticorrelation. Thus, we can exclude
that this is an artefact of the fitting technique used for determining
the peak magnitudes of the SNe. However, other systematic
uncertainties, which are not included in the error bars, 
could contribute to decrease the significance of this detection.  

In the right panel, we plot the scaled magnitude residuals, $X$, at
every point in the sky by taking the weighted average of the residual
for all SNe. The weight for each SNe is exponentially decreasing with
respect to the distance to the point where the average is computed,
using an exponential scale length of $15^\circ$. 

%%%%%%%%%%%%%%%%%%%%%%%%%%%%%%%%%%%%%%%%%%%%%%%%%%%%%%%%%%%%%%%%%%%%%%%
\section{Future data sets}\label{future}
%%%%%%%%%%%%%%%%%%%%%%%%%%%%%%%%%%%%%%%%%%%%%%%%%%%%%%%%%%%%%%%%%%%%%%%
As already noted, different SN surveys have very different survey
geometries. The survey geometry affects the distribution of the
angular separations which, in turn, affects on what scales and with
what precision it is possible to detect possible correlations. We now
consider the expected detection limit for SN~Ia magnitude residual angular
correlations for four different survey geometries, corresponding to
the complete SDSS-II, SNLS and SNAP surveys as well as an all-sky survey.

%======================================================================
\subsection{SDSS-II}
%======================================================================
The SDSS-II\footnote{\tt www.sdss.org} (Sloan Digital Sky Survey) Supernova Survey's 
survey area, called `Stripe 82' (southern equatorial stripe), is a region 2.5 degrees wide
along the celestial equator (i.e. $-1.25^{\circ} \le
\delta \le 1.25^{\circ}$), from roughly $-60^{\circ}<\alpha
<60^{\circ}$. They aim of the survey is to find roughly 500 SNe~Ia with $0.1\lesssim z \lesssim 0.3$ in this area out of which 300 SNe~Ia are expected to be used to build the Hubble diagram.

%======================================================================
\subsection{SNLS}
%======================================================================
The Supernova Legacy Survey\footnote{\tt www.cfht.hawaii.edu/SNLS} (SNLS) is 
a pencil-beam survey, using four small square patches, each covering a one-square-degree 
area (see figure~\ref{fig:snmaps} in section~\ref{datasets} for the patch configuration). 
This is an example of a survey geometry that limits the detection of possible correlations to certain angular scales. One can either look for correlations inside the patches or between patches (six different combinations). Around 500 SNe~Ia with redshifts up to $z \sim 1$ are expected to be used to build the final Hubble diagram.

%======================================================================
\subsection{SNAP}
%======================================================================
The survey geometry of the upcoming SNAP\footnote{\tt snap.lbl.gov} (SuperNova Acceleration
Probe) satellite is not yet determined, but the plan is to cover a
7.5-square-degree field in both the northern and southern hemisphere. In
total, SNAP is expected to find 2000 SNe~Ia with $z\lesssim 1.7$. We assume that the patches
have a square geometry.

%======================================================================
\subsection{All sky}
%======================================================================
Pan-STARRS\footnote{\tt pan-starrs.ifa.hawaii.edu} (Panoramic Survey Telescope and Rapid Response System) will as a by-product be able to detect a large number of SNe Ia across the entire sky.
The proposed LSST\footnote{\tt www.lsst.org} (Large Synoptic Survey Telescope) will search for faint astronomical objects across the entire sky and detect large amounts of SNe Ia. Depending on the observational effort put into following up these detections in the future, it is not unreasonable to assume that this could result in up to 2000 SNe Ia being put on the Hubble diagram at low and intermediate $z$ within the next decade.

%======================================================================
\subsection{Results}
\begin{figure}
\begin{center}
\includegraphics[angle=0,width=.48\textwidth]{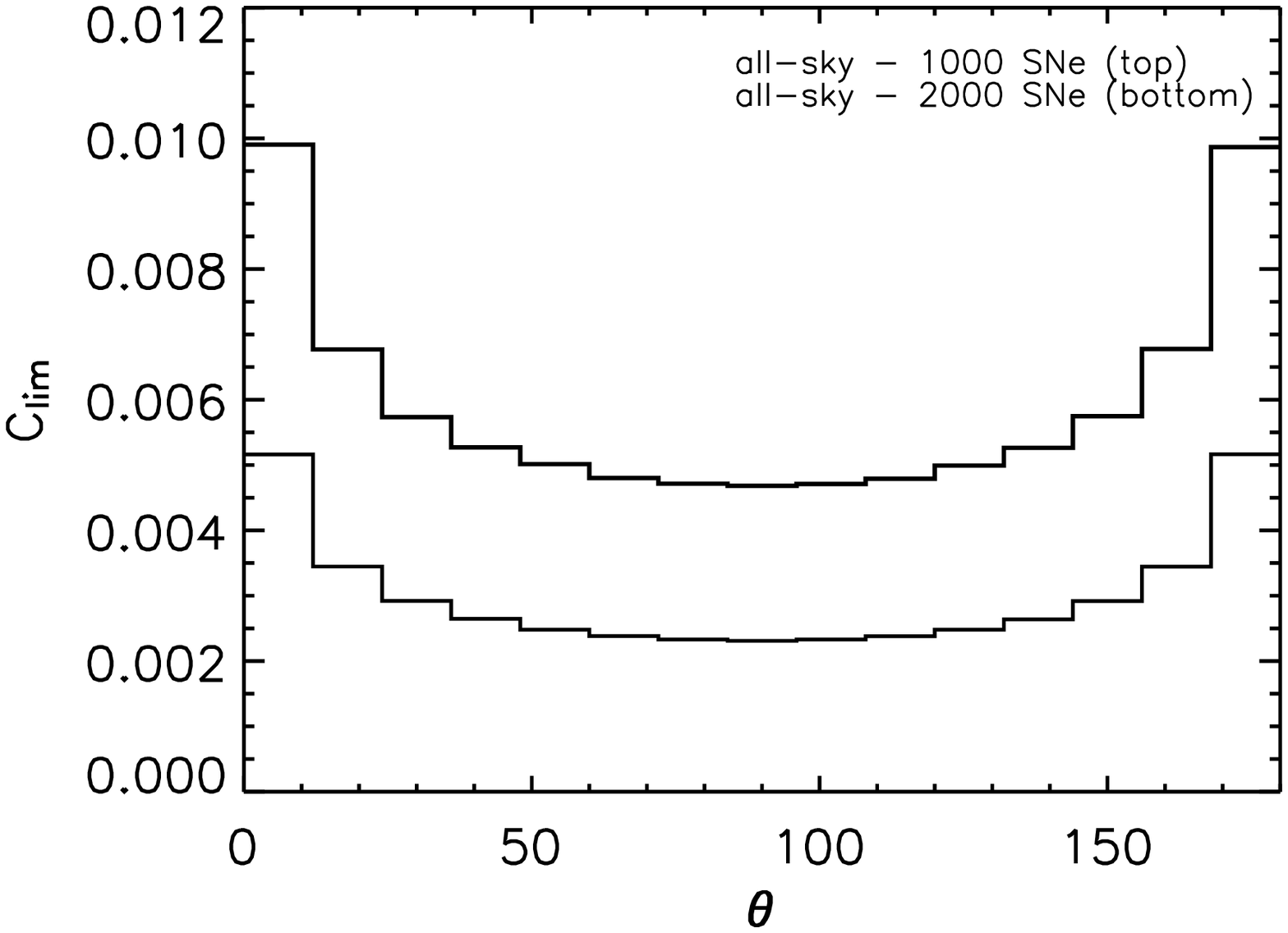}
\includegraphics[angle=0,width=.48\textwidth]{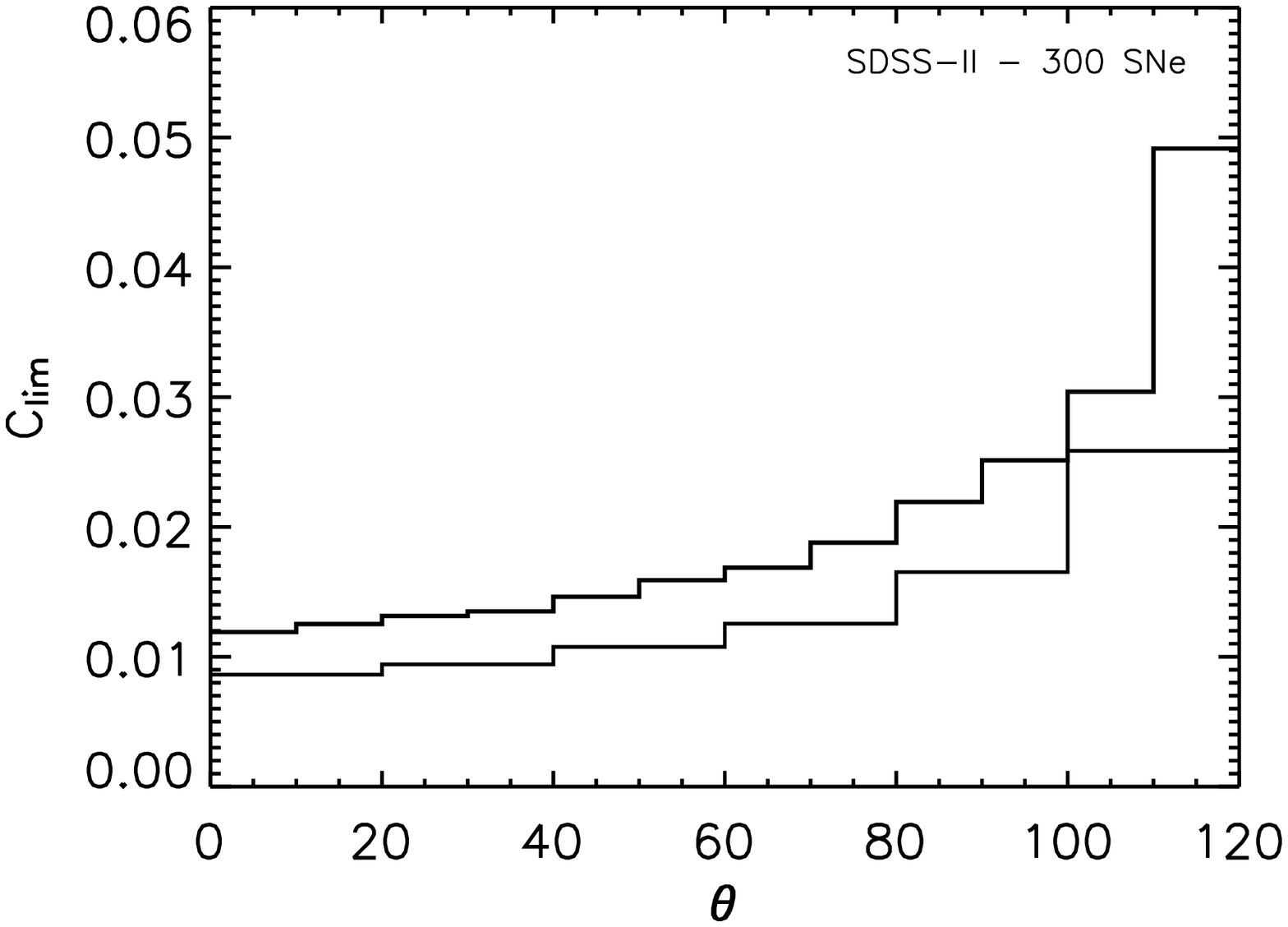}
\includegraphics[angle=0,width=.48\textwidth]{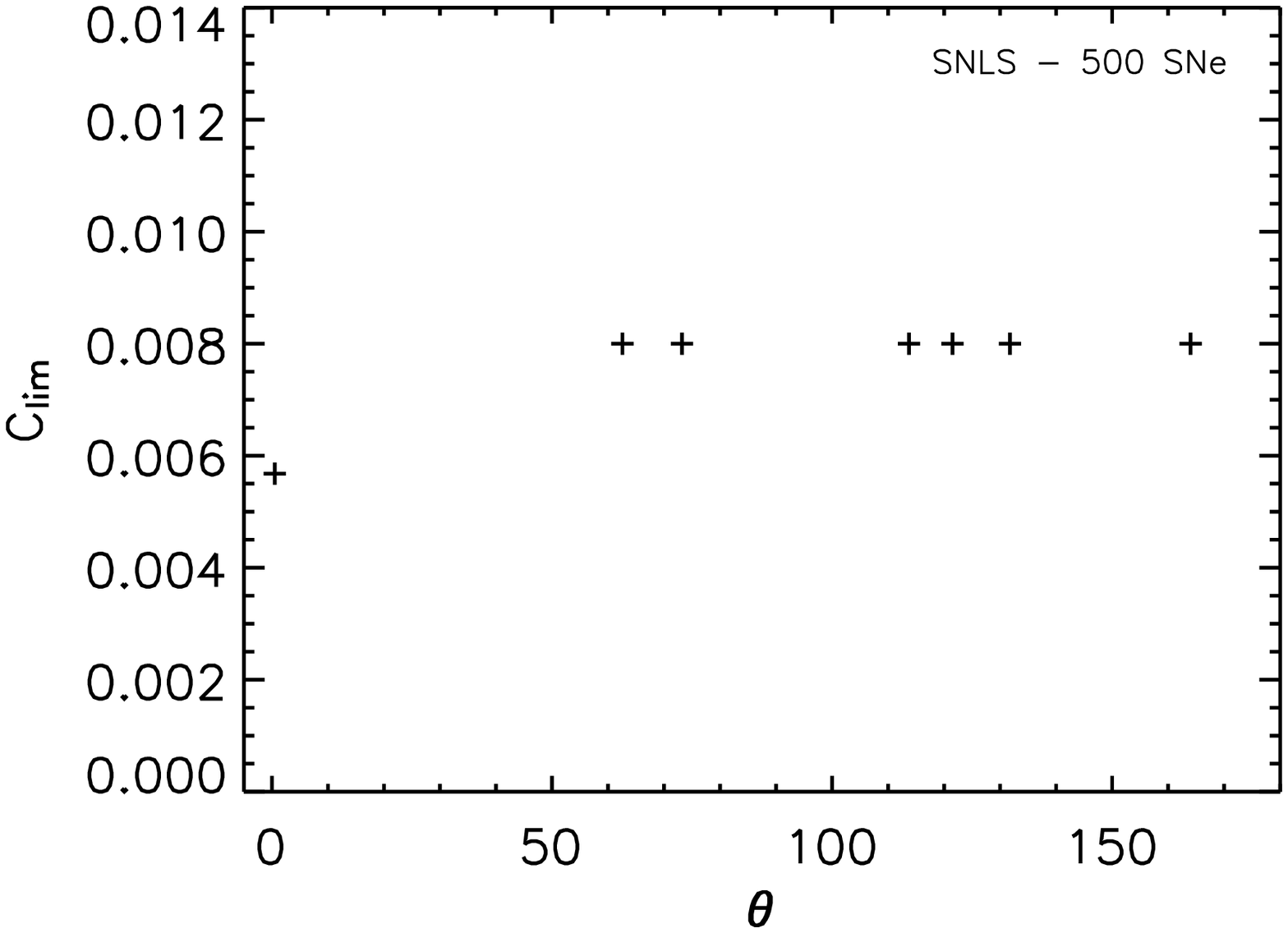}
\includegraphics[angle=0,width=.48\textwidth]{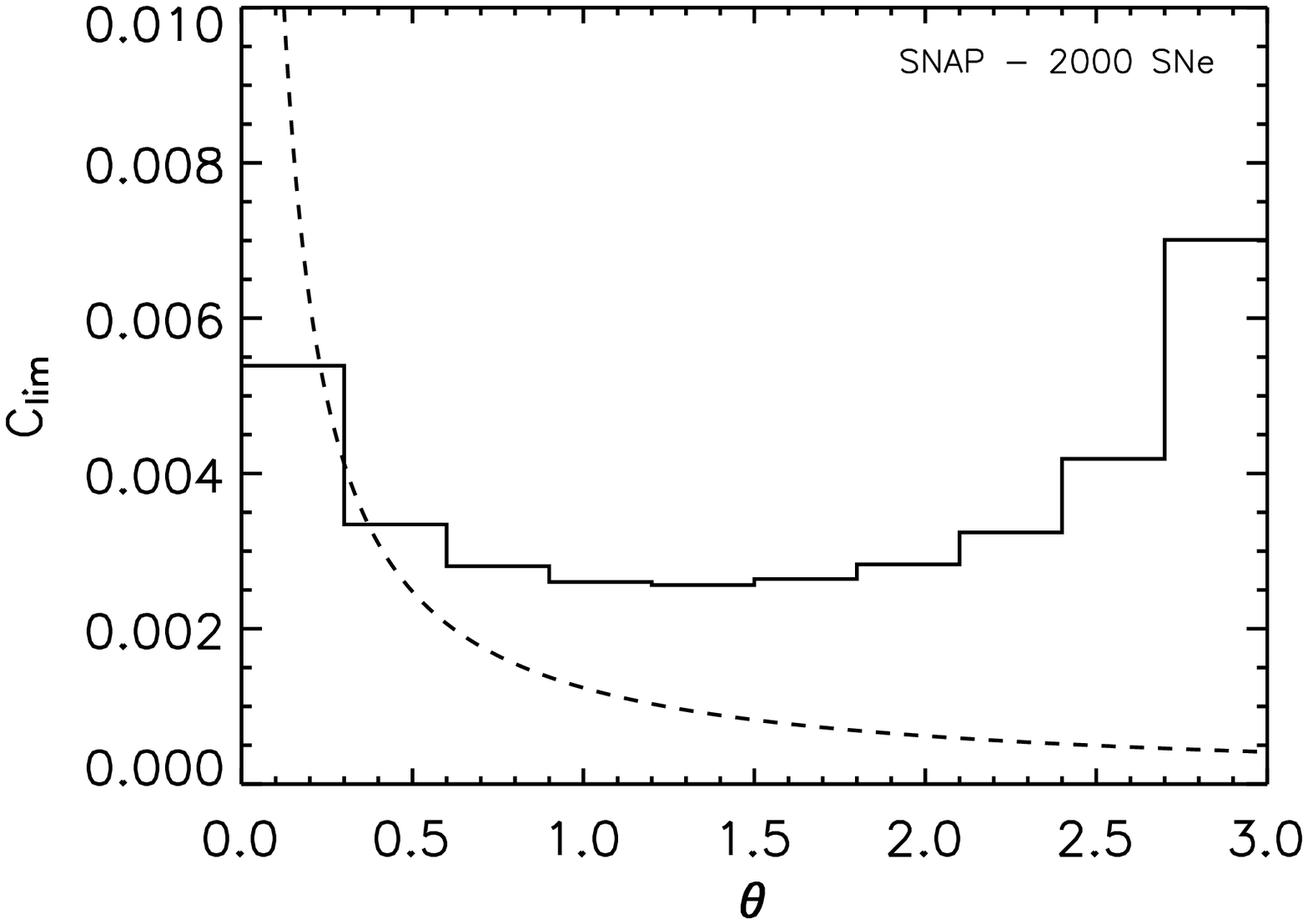}
\caption{\label{fig:corrlimits} 1-$\sigma$ detection limits for correlations in
the magnitude residuals in different surveys. Top left panel: all-sky
survey with 1000 SNe (top curve) and 2000 SNe (bottom curve). Top
right panel: SDSS-II with 300 SNe for two different bin sizes. Bottom
left panel: SNLS with 500 SNe. Note that the pencil-beam geometry only lets us probe certain discrete angular separations. Bottom right panel: SNAP with 2000 SNe for $\theta <3^{\circ}$. 
The overplotted dashed line is the lensing correlation of equation~(\ref{lenscorr}) with $z_{\rm s}=1$.}
\end{center}
\end{figure}

%======================================================================
We generate random SN positions in the ranges defined by the 
survey geometries and compute the 1-$\sigma$ detection limits $C_{\rm lim}$ for each of the four surveys described above. In the results presented in figure~\ref{fig:corrlimits}, we have assumed that the magnitude residuals in each bin correlate in the same way,
i.e., the error bars do not reflect the fact that the correlation
function may vary over the size of the bin. 

The amplitudes and detailed shapes of
the individual curves depend on the desired resolution. The lower the resolution, the larger the number of SN pairs in each bin and the lower the detection limit. We can scale
the curves by making the crude estimate that the number of SN pairs in
each bin scales with the bin size $\Delta \theta$ as $N_{\rm p}\propto
\Delta \theta$, such that $C_{\rm lim}$ decreases by a factor of
$\sqrt K$ if the resolution is degraded by a factor of $K$.

For the all-sky survey, there will be a peak at $90^{\circ}$ in the distribution of
the angular separations, with relatively few SN pairs at the
smallest and largest separations. The all-sky survey thus probes correlations around  
$\theta =90^{\circ}$ most effectively.
In the top left panel of figure~\ref{fig:corrlimits}, a resolution of $12^{\circ}$ is assumed. 
The top curve is for a data set with 1000 SNe and the bottom curve for 2000 SNe.  

For the SDSS-II, the angular separations are distributed with a maximum at
small angles and then fall off for increasing separations. 
Beyond $120^{\circ}$ it is not possible to probe possible correlations with this survey. 
In the top right panel of figure~\ref{fig:corrlimits}, the two curves are for a 
data set with 300 SNe, with an angular resolution in the top curve of $10^{\circ}$
and in the lower curve of $20^{\circ}$. The SDSS-II has its lowest detection 
limit at small angular scales.

Due to the pencil-beam geometry of the SNLS, it is only possible to probe certain discrete angular separations in this survey. We generate 125 SNe in each of the patches D1-D4 described in section~\ref{datasets}, for a total of 500 SNe. This means that there are equally many SN pairs in the bins, except for the first one, which probes correlations inside the patches. 
This bin contains more SN pairs and thus has the lowest detection limit as is evident from the lower left panel in figure~\ref{fig:corrlimits}.

For the SNAP survey, we generate 1000 SNe in each patch, for a total of 2000 SNe. The distribution of the angular separations inside the patches resembles the shape for the all-sky
survey but with a small tail at the high end. In figure~\ref{fig:corrlimits} we have cut off the part of the distribution with $3^{\circ}<\theta <\sqrt{15^{\circ}}$ for readability reasons. 

%======================================================================
\subsection{Lensing correlations}\label{lens}
%======================================================================
As noted in the introduction, physical effects such as peculiar motions and 
gravitational lensing will introduce correlations in 
the SN Ia data. Out of these, gravitational lensing will be the dominant contributor at 
$z \gtrsim 0.2$ \cite{2006PhRvD..73b3523B}.
The covariance of the weak lensing convergence $\kappa$ for two sources at the same
redshift $z_{\rm s}$ can be estimated using \cite{2002SSRv..100...73M}
\begin{equation}
\langle \kappa \cdot \kappa (\theta)\rangle^{1/2}\approx 0.01\sigma
_{8}\Omega _{\rm m}^{0.75}z_{\rm s}^{0.8}\Bigg(
\frac{\theta}{1^{\circ}} \Bigg)^{-(n+2)/2}\ ,
\end{equation}
where $n$ is the spectral index of the power spectrum of density
fluctuations. For small convergence ($\kappa \ll 1$) the magnitude
residual correlation due to weak lensing is given by
\begin{equation}\label{lenscorr}
\langle X\cdot X(\theta)\rangle\ \approx \Bigg(
\frac{-5}{\ln(10)\sigma _{m}} \Bigg)^2\langle \kappa \cdot \kappa
(\theta)\rangle \approx 10^{-3}\left(\frac{\sigma_m}{0.2}\right)^{-2}z_{\rm s}^{1.6}
\left(\frac{\theta}{1^{\circ}}\right)^{-1}\ ,
\end{equation}
where we have assumed $\sigma_8 = 0.8, \Omega _{\rm m}=0.3$ and $n=-1$. 
It is evident that correlations from gravitational lensing will only be 
important at small angular scales where we expect DE inhomogeneities to be negligible.
A tentative detection of the effects of gravitational lensing on SNe~Ia in the 
GOODS fields was made in J\"onsson \textit{et al} \cite{2007JCAP...06..002J} by
cross-correlating the SN~Ia residuals with the lensing properties of foreground galaxies.
The dense sampling of the SNAP survey fields will allow us to detect the angular 
correlation of SN~Ia residuals without using the information from the lensing galaxies.
In the bottom right panel of figure~\ref{fig:corrlimits}, the lensing correlation 
for a source redshift of $z_{\rm s}=1$ (approximately equal to the average SN redshift expected for SNAP) and $\sigma _m = 0.2$ is compared with the SNAP correlation detection.
It is clear that SNAP will be able to see the lensing correlation with a resolution of $<0.3^\circ$; see also \cite{2006ApJ...637L..77C} for how this information can be used to probe cosmological parameters.
We also note that with the increased statistics, it will be possible to probe the angular correlation function in separate redshift intervals, or to do a full three-dimensional correlation analysis.   

%%%%%%%%%%%%%%%%%%%%%%%%%%%%%%%%%%%%%%%%%%%%%%%%%%%%%%%%%%%%%%%%%%%%%%%
\section{Toy model}\label{toymodel}
%%%%%%%%%%%%%%%%%%%%%%%%%%%%%%%%%%%%%%%%%%%%%%%%%%%%%%%%%%%%%%%%%%%%%%%
In this section, we assume a toy model angular correlation function, and
illustrate how well we can detect such a correlation using future data
expected for the SDSS-II and SNLS surveys described in
section~\ref{future}. We let the magnitude residuals be the sum of an
uncorrelated part, $X_{\rm i}$, and a correlated part, $X_{\rm c}$, due
to, e.g., inhomogeneous DE, 
\begin{equation}
X=X_{\rm i}+X_{\rm c}\ .
\end{equation} 
Our toy correlation function has the form of a Gaussian,
\begin{equation}\label{corrfunc}
C(\theta)=A\exp\{-[(\theta -\theta _{\rm a})/\theta _{\rm c}]^2\}\ ,
\end{equation}
where the amplitude, $A$, is the variance of the correlated magnitude
residuals, $\sigma _{X_{\rm c}}^{2}=A$, $\theta _{\rm a}$ is the angle
where the correlation has its maximum value, and $\theta _{\rm c}$ is
the angular scale of the variation in the correlation function.

\begin{figure}
\begin{center}
\includegraphics[angle=0,width=.48\textwidth]{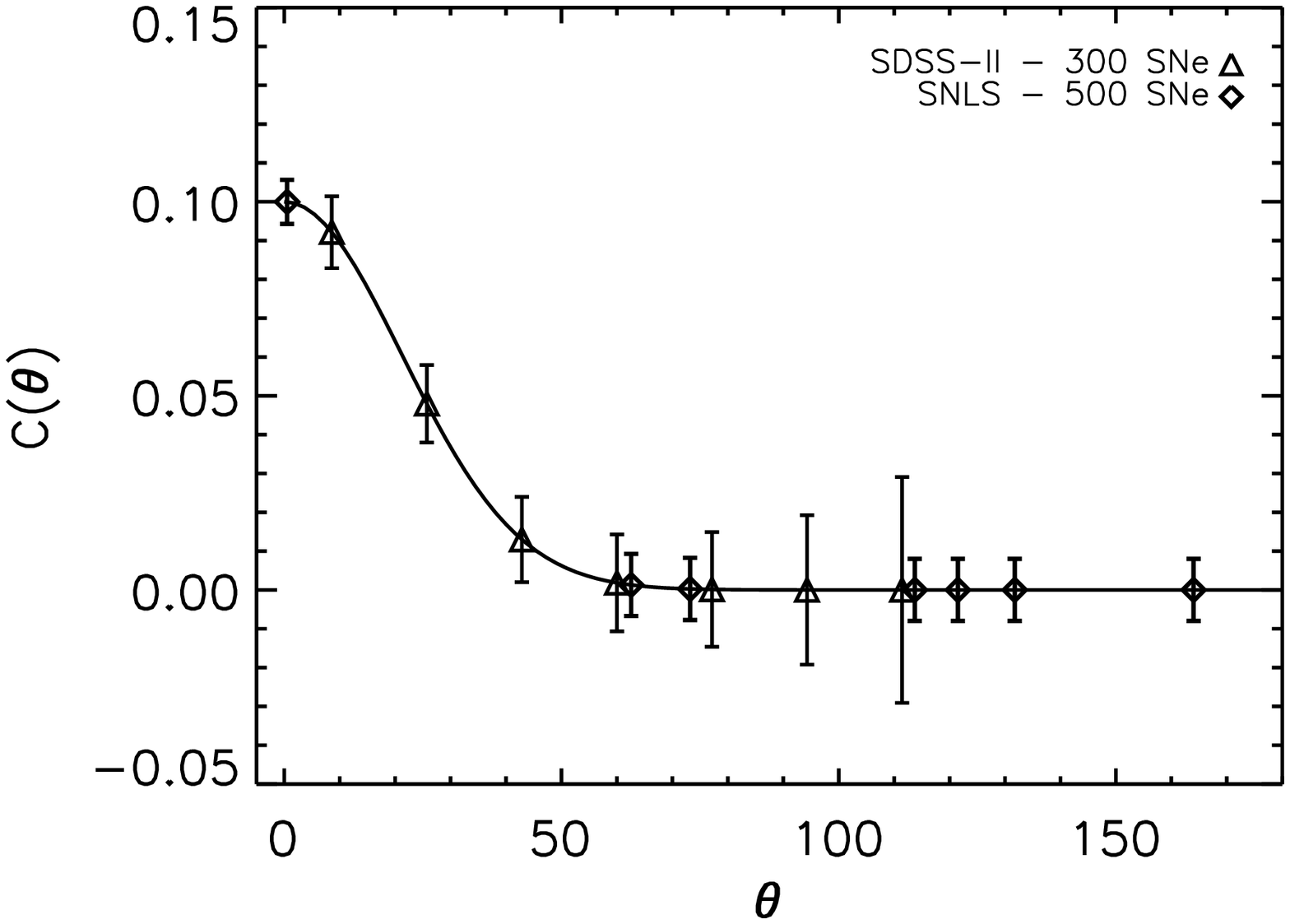}
\includegraphics[angle=0,width=.48\textwidth]{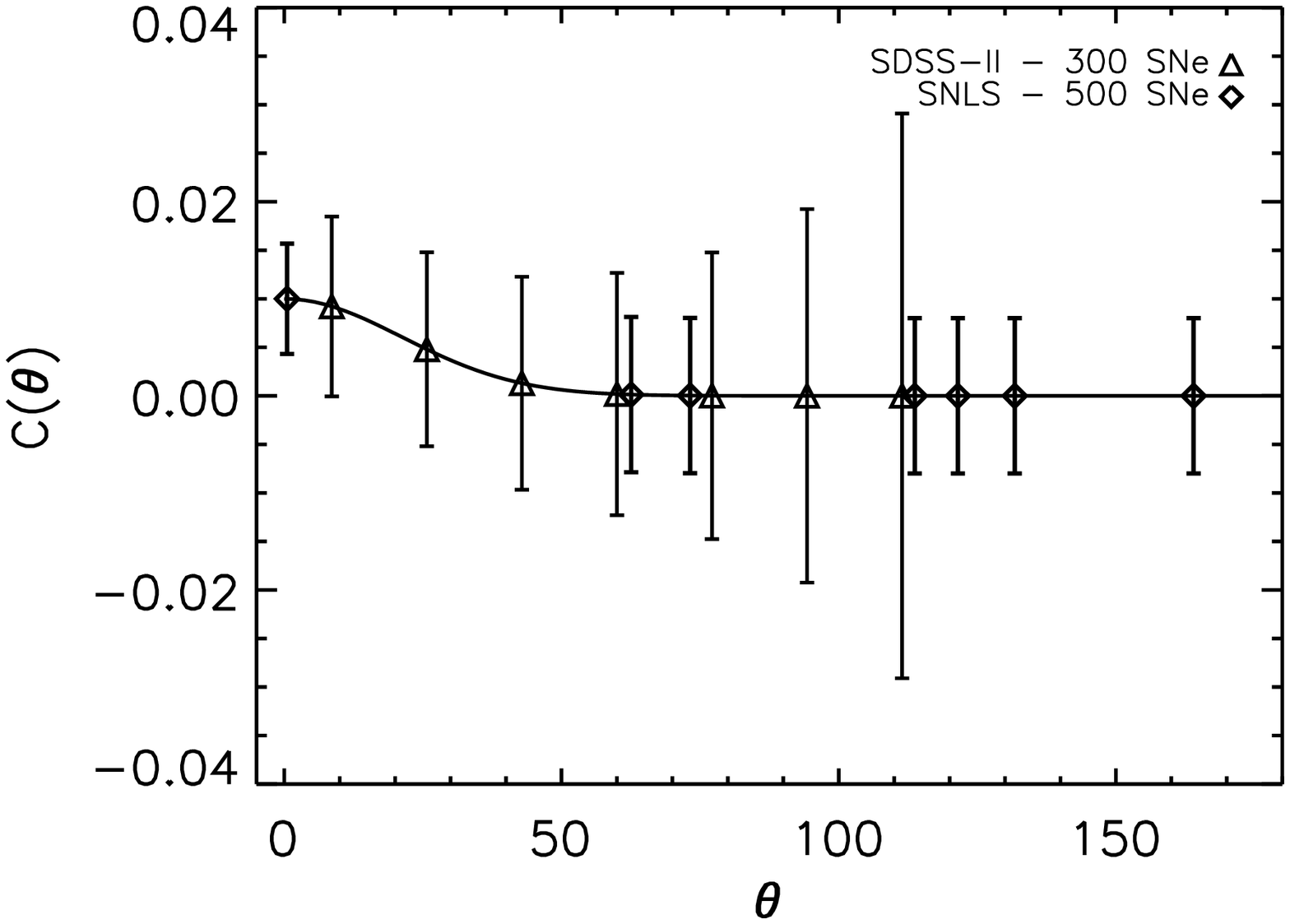}
\caption{\label{fig:SDSSSNLS} Our toy correlation function with $\theta
_{\rm c}=30^{\circ}$ and $\theta _{\rm a}=0^{\circ}$ for 300 SDSS-II SNe (triangles) and 500 SNLS 
SNe (diamonds). The error bars represent the 68\% confidence limit. Left panel: correlation function amplitude $A=0.1$, corresponding to $\sigma _{X_{\rm i}}/\sigma _{X_{\rm
c}}=3$. Right panel: $A=0.01$, corresponding to $\sigma _{X_{\rm
i}}/\sigma _{X_{\rm c}}\approx 10$. The SDSS-II has uniform binning while the SNLS correlation is evaluated at the seven discrete angular separations defined by the survey geometry.}
\end{center}
\end{figure}

We generate random SN positions and uncorrelated intrinsic magnitude
residuals with a mean of zero and a variance of $\sigma _{X_{\rm
i}}^{2}=1-A$. The ratio of the dispersions of the
intrinsic and the correlated magnitude residuals is
\begin{equation}
\frac{\sigma _{X_{\rm i}}}{\sigma _{X_{\rm c}}}=\sqrt{\frac{1-A}{A}}\ .
\end{equation}
Figure \ref{fig:SDSSSNLS} shows the correlation function
equation~(\ref{corrfunc}) with $\theta _{\rm c}=30^{\circ}$ and $\theta _{\rm a}=0^{\circ}$. 
The error bars represent the 68\% confidence limit obtained from equation~(\ref{clim}) 
for 300 SDSS-II SNe (triangles) and 500 SNLS SNe (diamonds). We have neglected the fact that the correlation function varies over the range of each bin.  
The binning is uniform for the SDSS-II. For the SNLS, we have used the discrete angular separations defined by the survey geometry. 
The left panel shows the correlation function with $A=0.1$, corresponding to $\sigma _{X_{\rm i}}/\sigma _{X_{\rm c}}=3$. 
The correlation function is detected at high confidence using either data set. The right panel shows the correlation
function with $A=0.01$, corresponding to $\sigma _{X_{\rm i}}/\sigma _{X_{\rm c}}\approx 10$. 
The SDSS-II data set is not sufficient for detecting the correlation in this case. For the SNLS, we expect to be able to 
make a weak detection of the correlation within the four survey patches. 

%%%%%%%%%%%%%%%%%%%%%%%%%%%%%%%%%%%%%%%%%%%%%%%%%%%%%%%%%%%%%%%%%%%%%%%
\section{Conclusions}\label{conclusions}
%%%%%%%%%%%%%%%%%%%%%%%%%%%%%%%%%%%%%%%%%%%%%%%%%%%%%%%%%%%%%%%%%%%%%%%

The cosmological community is hard at work trying to constrain the
behaviour of DE. The most pressing question is whether DE is a
CC or something dynamical. A detection of temporal
or spatial variations of DE would answer this question and
refute the assertion of the CC being the dominant energy component in the universe. At the
same time, there are alternative attempts to explain the apparent
acceleration of the universe without invoking DE at all:
instead as originating in the large scale inhomogeneities in the matter distribution. 
Irrespective of their origin, large scale inhomogeneities would
manifest themselves as anisotropies in the observed magnitudes of SNe~Ia. 

In this paper, we have devised a methodology for detecting angular
correlations in SN~Ia magnitude residuals. The methodology was applied
to two recently found data sets (astier06 and davis07), neither of which show
any signs of correlations at angular scales 
$0^\circ<\theta<180^\circ$. The uncertainties on the measured
correlations, $C(\theta)$, are approximately 10\% and 5\% for the
astier06 and davis07 data sets respectively, using an angular
resolution of $\sim 15^\circ$.  Note, however, that the two data sets are
not independent, since they overlap partially for the high redshift
sample, and include almost the same set of nearby SNe.  
The main difference between the data sets is that different
assumptions on SN~Ia properties, such as intrinsic colour and dust
extinction in the host galaxy, are used when deriving the SN~Ia peak
magnitudes. Such assumptions are responsible for some of the most
important systematic uncertainties in SN~Ia cosmology. A comparison
between the results for the two data sets allows us to keep the impact
of these uncertainties on the measured correlation function under control.  
Because systematic uncertainties are typically different at low and
high redshifts, and we expect effects from DE inhomogeneities to grow with redshift, 
we have applied the same analysis to the two sub-samples
separately. At low $z$, we found an anticorrelation at the 2$\sigma$ level
at angular scales $\theta \approx 40^{\circ}$. Due to the weak
statistical significance of this detection, and the low redshifts, 
we cannot draw any conclusions about the possible implications for DE
inhomogeneity. It is worth noticing that the nearby SNe analysed are a
collection of very inhomogeneous data samples, collected by different
observers at various telescopes and thus, there could be systematic
uncertainties involved which have not been included in the error bars
(see, e.g., discussion in \cite{2003A&A...404..901N}). 
We note that we see no signs of anisotropy on angular scales
$\theta \sim 180^{\circ}$, corresponding to the north/south
asymmetry detected in \cite{2007A&A...474..717S}. Our
data sample, however, is only partially overlapping with that work, 
since we have only included SNe in the Hubble flow ($z \geq 0.015$)
in our analysis in order to mitigate the effects from 
correlations arising from peculiar motions.

Gravitational lensing correlations will only be
significant on small scales, where we expect the effects from DE
clustering to be negligible. Using data from the proposed SNAP
satellite, we should be able to detect the lensing correlations in SN~Ia
magnitudes with an angular resolution of $<0.3^\circ$. 

Correlations induced by intervening galactic dust will also mainly
take place on very small angular scales, whereas correlations from
dust in the Milky Way could potentially be a problem for an all-sky
survey. Most surveys, however, observe in directions in the sky where
the Milky Way dust extinction is well measured and understood.  

More SN data are needed if we are to detect
presumably weak correlations and better constrain
inhomogeneous/anisotropic models. Our simulations of future data 
illustrate how the survey geometries govern on what scales, and to
what precision, we will be able to detect possible angular
correlations. We find that using data 
from the soon to be completed SDSS-II and SNLS surveys, we will
typically be able to detect correlations in the magnitude residuals at
the per cent level. 

Any claim of a deviation of DE properties from those of a CC will be
subject to  intense scrutiny from the cosmological community, and will
need to be backed up by independent evidence in order to be generally
accepted. It is therefore of utmost importance to pursue the study of
DE along multiple paths. Although observationally demanding, the
search for spatial variations in DE properties is a useful
complement to studies aiming at constraining the time evolution of
DE. Specifically, for SNe~Ia, we expect systematic effects connected
to the time evolution of DE, such as temporal evolution of SN~Ia
properties, to be of less importance when studying spatial variations
of DE. Since systematic effects constitute the limiting factor for
future SN~Ia surveys, we expect interesting results for the spatial
correlations of SN magnitudes in the near future, irrespective of
whether such correlations are found or not.    

%%%%%%%%%%%%%%%%%%%%%%%%%%%%%%%%%%%%%%%%%%%%%%%%%%%%%%%%%%%%%%%%%%%%%%%
\ack
%%%%%%%%%%%%%%%%%%%%%%%%%%%%%%%%%%%%%%%%%%%%%%%%%%%%%%%%%%%%%%%%%%%%%%%
MB acknowledges support from the HEAC Centre funded by the Swedish Research Council. EM acknowledges support for this study by the Swedish Research Council
and from the Anna-Greta and Holger Crafoord fund. The authors would like to thank
Ariel Goobar for useful discussions.

%%%%%%%%%%%%%%%%%%%%%%%%%%%%%%%%%%%%%%%%%%%%%%%%%%%%%%%%%%%%%%%%%%%%%%%
\section*{References}
%%%%%%%%%%%%%%%%%%%%%%%%%%%%%%%%%%%%%%%%%%%%%%%%%%%%%%%%%%%%%%%%%%%%%%%

\end{document}